# Spin-transfer torque effects in the dynamic forced response of the magnetization of nanoscale ferromagnets in superimposed ac and dc bias fields in the presence of thermal agitation

D. J. Byrne,[1] W. T. Coffey,[2] Y. P. Kalmykov,[3] S. V. Titov,[4] and J. E. Wegrowe[5]

[1]School of Physics, University College Dublin, Belfield, Dublin 4, Ireland

[2]Department of Electronic and Electrical Engineering, Trinity College, Dublin 2, Ireland

[3]Laboratoire de Mathématiques et Physique, Université de Perpignan Via Domitia, F-66860, Perpignan, France

[4]Kotel'nikov Institute of Radio Engineering and Electronics of the Russian Academy of Sciences, Vvedenskii Square 1, Fryazino, Moscow Region, 141120, Russia

[5]Laboratoire des Solides Irradiés, Ecole Polytechnique, 91128 Palaiseau Cedex, France

**Abstract**

Spin-transfer torque (STT) effects on the stationary forced response of nanoscale ferromagnets subject to thermal fluctuations and driven by an ac magnetic field of arbitrary strength and direction are investigated via a generic nanopillar model of a spin-torque device comprising two ferromagnetic strata representing the free and fixed layers and a nonmagnetic conducting spacer all sandwiched between two ohmic contacts. The STT effects are treated via Brown's magnetic Langevin equation generalized to include the Slonczewski STT term thereby extending the statistical moment method [Y. P. Kalmykov *et al.*, Phys. Rev. B **88**, 144406 (2013)] to the forced response of the most general version of the nanopillar model. The dynamic susceptibility, nonlinear frequency-dependent dc magnetization, dynamic magnetic hysteresis loops, etc. are then evaluated highlighting STT effects on both the low-frequency thermal relaxation processes and the high-frequency ferromagnetic resonance, etc., demonstrating a pronounced dependence of these on the spin polarization current and facilitating interpretation of STT experiments.

# I. INTRODUCTION

One of the most significant developments in magnetization reversal by thermal agitation in nanoscale ferromagnets since the seminal treatment of Néel [1] and Brown [2] has been the spin-transfer torque (STT) effect [3-5] existing because an electric current with spin polarization in a ferromagnet has an associated flow of angular momentum [3-7] thereby exerting a macroscopic spin-torque. Consequently, the magnetization **M** of the ferromagnet *may be altered by spin-polarized currents*, which underpin the novel subject of spintronics [8], i.e., current-induced control over magnetic nanostructures. Applications include (a) very high speed current-induced magnetization switching by reversing the orientation of magnetic bits [5,9] and (b) using spin-polarized currents to manipulate steady state microwave oscillations [9] via the steady state magnetization precession due to STT representing the conversion of dc input current into an ac output voltage [5]. Now due to thermal fluctuations [5,9], STT devices invariably represent an *open* system on the *nanoscale* in an out-of-equilibrium steady state quite unlike conventional nanostructures characterized by the Boltzmann equilibrium distribution. Therefore, the thermal fluctuations cannot be ignored. They lead to mainly noise-induced switching at currents far less than the critical switching current without noise as well as introducing randomness into the precessional orbits, which now exhibit energy-controlled diffusion [10]. Thus, the effect of the noise is generally to reduce the current-induced switching time. This phenomenon has been corroborated by many experiments (e.g., [11]) demonstrating that STT near room temperature alters thermally activated switching processes, which then exhibit a pronounced dependence on both material and geometrical parameters. However, in marked contrast to the well-developed zero temperature limit, $T = 0$, and to nanomagnets at *finite* temperature without STT, various treatments of the thermally activated magnetization reversal in STT systems (e.g., escape rates [12-15] and stochastic dynamic simulations [16-19]) are still in a state of flux [20]. Therefore, accurate solutions of generic STT models at finite temperatures are necessary both to properly assess such theories and to achieve further improvements in the design and interpretation of experiments, particularly due to the manifold practical applications in spintronics, random access memory technology, and so on.

The archetypal model (Fig. 1) of a STT device is a nanostructure comprising two magnetic strata labelled the *free* and *fixed* layers and a nonmagnetic conducting spacer. The fixed layer is much more strongly pinned along its orientation than the free one. On passing an electric current through the fixed layer it becomes spin-polarized which, as it encounters the free layer, induces a STT that alters the magnetization **M** of that layer. Both ferromagnetic layers are assumed to be uniformly magnetized [8]. Although the single-domain or macrospin approximation cannot explain all observations of the magnetization dynamics in spin-torque systems, nevertheless many qualitative features needed to interpret experimental data are satisfactorily reproduced. Thus the current-induced magnetization dynamics in the free layer including thermal fluctuations may be described by the Landau-Lifshitz-Gilbert-Slonczewski equation [3], i.e., the Landau-Lifshitz-Gilbert equation [21] including the STT augmented by a Gaussian white noise field $\mathbf{h}(t)$ so becoming a Langevin equation [5,7,20], viz.,

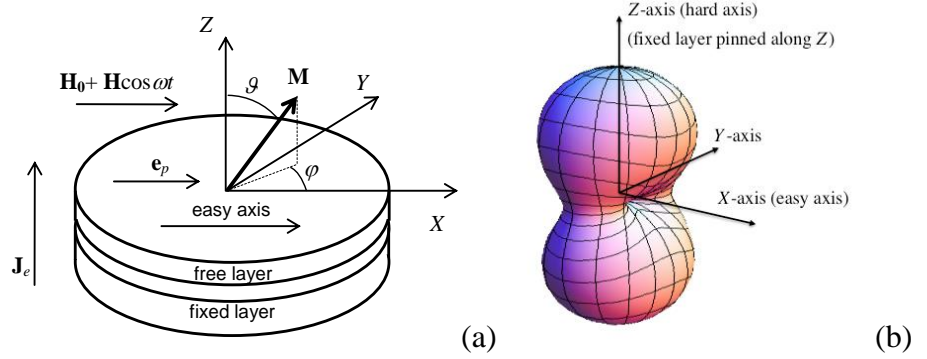

FIG. 1. (Color on line) (a) Geometry of the problem: A STT device consists of two ferromagnetic strata labelled the *free* and *fixed* layers, respectively, and a normal conducting spacer all sandwiched on a pillar between two ohmic contacts [12]. The fixed layer has a fixed magnetization along the direction $\mathbf{e}_P$. $\mathbf{J}_e$ is the spin-polarized current density, $\mathbf{M}$ is the magnetization of the free layer, $\mathbf{H}_0$ is the dc bias magnetic field and $\mathbf{H}\cos\omega t$ is the applied ac field. (b) Free energy potential presented in the standard form of superimposed easy-plane and in-plane easy-axis anisotropies.

$$\dot{\mathbf{u}} = -\gamma \mathbf{u} \times \left(\mathbf{H}_{\text{eff}} + \dot{\mathbf{u}}\big|_{\text{ST}} + \mathbf{h}\right) + \alpha \mathbf{u} \times \dot{\mathbf{u}}. \tag{1}$$

Here $\mathbf{u} = M_S^{-1}\mathbf{M}$ is a unit vector along $\mathbf{M}$, $M_S$ is the saturation magnetization, $\gamma$ is the gyromagnetic-type constant, $\alpha$ is a dimensionless phenomenological damping parameter, representing the combined effect of all the microscopic degrees of freedom,

$$\mathbf{H}_{\text{eff}} = -\frac{1}{\mu_0 M_S}\frac{\partial V}{\partial \mathbf{u}} \tag{2}$$

is the effective magnetic field comprising the anisotropy and external applied fields, while $\partial/\partial\mathbf{u}$ denotes the gradient operator on the surface of the unit sphere, $\mu_0 = 4\pi \cdot 10^{-7}\,\text{JA}^{-2}\text{m}^{-1}$ in SI units, $V$ is the free energy density of the free layer, and the STT term $\dot{\mathbf{u}}\big|_{\text{ST}}$ in Eq. (1) is defined as

$$\dot{\mathbf{u}}\big|_{\text{ST}} = -\frac{1}{\mu_0 M_S}\mathbf{u}\times\frac{\partial \Phi}{\partial \mathbf{u}},$$

where $\Phi$ is the non-conservative potential due to the spin-polarized current [3,4,20].

Almost invariably, the effects of thermal fluctuations combined with STT have been investigated via the magnetic Langevin equation (1) or its associated Fokker-Planck equation [20]. The magnetic Langevin equation without STT was originally proposed by Brown [2] for theoretical treatment of the magnetization reversal in magnetic nanoparticles. His primary objective was to securely anchor Néel's conjectures [1] concerning the nature of the superparamagnetic relaxation of a single domain ferromagnetic particle within the framework of the theory of stochastic processes (in essence, Brown's theory of the magnetization relaxation in magnetic nanoparticles [2] is an analog of the Debye theory [22,23] of dielectric relaxation of polar liquids). During the last decade, various analytical and numerical approaches to the calculation of the measurable parameters of STT devices via the magnetic Langevin equation including STT have been developed. These include generalizations (e.g., Refs. [12-14, 20]) of the Kramers escape rate theory [24-27] and stochastic dynamics simulations (e.g., Refs. 12, 16-19). For example, the pronounced time separation between *fast* precessional and *slow* energy changes in *lightly* damped ($\alpha \ll 1$) closed phase space trajectories (called Stoner-Wohlfarth orbits) at energies near the barrier energy has been exploited in Refs. [7, 12, 13] to formulate a one-dimensional Fokker-Planck equation for the energy distribution function essentially similar to that derived by Kramers [24] for point particles. These generalizations yield STT effects in the thermally assisted magnetization reversal via the Langevin and/or Fokker-Planck equations as a function of temperature, damping, external magnetic field, and spin-polarized current. In particular, varying the spin-polarized current may alter the reversal time by several orders of magnitude concurring with experimental results [11].

Now we have previously treated [28] STT effects on certain out-of-equilibrium time- and frequency-independent stationary observables in the presence of a dc *bias field alone* via the generic nanopillar model (Fig. 1) by solving the magnetic Langevin equation (1) using the statistical moment method [27]. These observables comprise the stationary distribution of the magnetization orientations, the effective potential, the in-plane component of the magnetization of the free layer, and the static susceptibility. In particular, these *time- and frequency-independent* observables have been studied [28] for wide ranges of the spin-polarized current, the dissipative coupling to the thermal bath, the anisotropy parameters and the magnitude and orientation of the applied external dc field, which was supposed *constant* in time. Besides the calculation of these stationary observables, the reversal time of the in-plane component of the magnetization of the free layer has also been evaluated [28] via the smallest nonvanishing eigenvalue of the corresponding Fokker-Planck operator [29] again as a function of the parameters mentioned. Now in Ref. 28, the external applied (bias) field was supposed *time-independent*, i.e., it represents a dc field applied in the infinite past. Hence, the results of Ref. 28 cannot be applied to virtually all *dynamical* aspects of the time-dependent magnetization response. These include magnetization switching of STT devices and line shape of STT nano-oscillators driven by ac external magnetic fields and currents [30-35], stochastic resonance [36-39], etc. In particular, as shown experimentally, the magnetization reversal in STT devices driven by superimposed dc and ac currents or by a direct spin-polarized current combined with an ac magnetic field may allow one a more efficient STT magnetization reversal in comparison to that achievable by purely dc currents alone. There the problem is of technological interest in the context of improving switching characteristics of magnetic random access memories [35]. Despite the potential applications, an accurate theoretical description of STT effects in the response of a nanomagnet to an ac force of *arbitrary* strength in the presence of thermal agitation has not yet been fully developed due to the inherent difficulties generally associated with modelling a nonlinear response. As a result, most of the theoretical methods, which were developed for STT effects (see, e.g., [30-39]), concern the ac response over *limited* ranges of the frequency and amplitude. Hence, they do not cover many other dynamical characteristics of nanomagnets including the nonlinear complex magnetic susceptibility and dynamic magnetic hysteresis (DMH) loops, which require the response to a strong ac magnetic field over a *wide* frequency range. Therefore, to comprehensively investigate the influence of STT on the dynamical characteristics of the generic nanopillar model (Fig. 1) due to an *ac magnetic field of arbitrary strength and frequency*, we generalize the approach of Titov *et al.* [40] developed originally for zero STT. The advantage of this approach over all others is that one can obtain the nonlinear response characteristics for all frequencies of interest ranging from the very low ones corresponding to overbarrier relaxation processes up to the very high frequencies appropriate to the ferromagnetic resonance (GHz) range using a single model. Now *a priori* STT effects in the ac stationary response of a nanomagnet inherently pose a more complicated problem than the time-independent out-of-equilibrium case of Ref. 28 because the observables are now both *time-* and *frequency-dependent*. However, these difficulties may be overcome using the matrix continued fraction method [27,29] just as with the nonlinear ac response without STT [40,41].

The paper is arranged as follows. In Sec. II, the basic equations for the calculation of the ac stationary response are given. In Sec. III, the spectra of the linear dynamic susceptibility in *all* frequency ranges characterizing the magnetization dynamics are given demonstrating a strong dependence on STT. In Sec. IV, STT effects on spectra of the nonlinear dynamic susceptibility and the stationary *time-independent* but *frequency-dependent* magnetization are illustrated, while STT effects on DMH loops and specific absorption rate are studied in Sec. V. Appendixes A and B contain a detailed account of the matrix continued fraction solution for the stationary response of a nanoscale ferromagnet to an ac magnetic field of arbitrary strength.

## II. STATISTICAL MOMENT EQUATIONS

Now the main thrust of our investigation is the study of STT effects on the complex magnetic susceptibility and DMH loops of a nanoscale ferromagnet subjected to superimposed ac and dc bias fields $\mathbf{H}_0 + \mathbf{H}\cos\omega t$ of arbitrary strengths and orientations using the generic nanopillar model illustrated by Fig. 1. Here the normalized free energy per unit volume $\beta V(\vartheta,\varphi,t)$ of the free layer may conveniently be written as ($\mathbf{H}_0$ and $\mathbf{H}$ are assumed *parallel*)

$$\beta V(\vartheta,\varphi,t) = \sigma\left(\delta\cos^2\vartheta - \sin^2\vartheta\cos^2\varphi\right) \\ -\left(\xi_0 + \xi\cos\omega t\right)\cos\Theta(\vartheta,\varphi), \quad (3)$$

where $\vartheta$ and $\varphi$ are the angular coordinates specifying the orientation of the magnetization $\mathbf{M}$ in spherical polar coordinates (see Fig. 1b), $\sigma = \beta\mu_0 M_S^2 D_\parallel$ and $\delta = D_\perp / D_\parallel$ are the dimensionless anisotropy and biaxiality parameters respectively, $D_\parallel$ and $D_\perp$ account for both demagnetizing and magnetocrystalline anisotropy effects [20], $\xi_0 = \beta\mu_0 M_S H_0$ and $\xi = \beta\mu_0 M_S H$ are the dc and ac external field parameters, respectively, $\beta = v/(kT)$, $v$ is the volume of the free layer, $k$ is Boltzmann's constant, $T$ is the absolute temperature, while $\Theta$ is the angle between **H** and **M** so that

$$\cos\Theta(\vartheta,\varphi) = (\mathbf{u}\cdot\mathbf{H})/H \qquad (4)$$
$$= \gamma_1 \sin\vartheta\cos\varphi + \gamma_2 \sin\vartheta\sin\varphi + \gamma_3 \cos\vartheta.$$

Here $\gamma_1 = \cos\varphi_\xi \sin\vartheta_\xi$, $\gamma_2 = \sin\varphi_\xi \sin\vartheta_\xi$, and $\gamma_3 = \cos\vartheta_\xi$ are the direction cosines of the applied dc and ac fields. The first term on the right hand side of Eq. (3), namely, $\sigma(\delta\cos^2\vartheta - \sin^2\vartheta\cos^2\varphi)$ constitutes a conservative potential taken in the standard form of superimposed easy-plane and in-plane easy-axis anisotropies (see Fig. 1b). This potential, in general, represents an energyscape with two minima and two saddle points compelling the magnetization to align in a given direction in either of the energy minima in the equatorial or *XY* plane [28]. As in Ref. [28], *Z* is taken as the hard axis while the *X*-axis is the easy one. Furthermore, the non-conservative potential $\Phi$ due to the spin-polarized current may sensibly be approximated [28] for all polar angles $\vartheta$, $\varphi$ and arbitrary orientation of the unit vector $\mathbf{e}_P$ (identifying the magnetization direction in the fixed layer) by a finite series of spherical harmonics $Y_{lm}(\vartheta,\varphi)$ [42], viz.,

$$\beta\Phi \cong \sum_{r=0}^{2}\sum_{s=-r}^{r} B_{rs} Y_{rs}(\vartheta,\varphi), \qquad (5)$$

where the expansion coefficients $B_{rs}$ are listed explicitly in Ref. [28].

Now the task of calculating the ac stationary response from the Langevin equation (1) can always be reduced to the solution of an infinite hierarchy of differential-recurrence relations for the statistical moments (averaged spherical harmonics $\langle Y_{lm}\rangle(t)$, where the angular brackets $\langle\ \rangle$ mean statistical averaging) as with zero STT [40,41]. Such a hierarchy has been derived in Ref. [28] for the non-conservative potential due to spin-polarized current given by Eq. (5) and the biaxial anisotropy plus the Zeeman term due to a spatially uniform dc bias field $\mathbf{H}_0$. In like manner, we can generalize this derivation to our case representing the response to a dc bias field temporally modulated by an ac field $\mathbf{H}\cos\omega t$. Here the total free energy density *V* is given by Eq. (3) above and the infinite hierarchy of 25-term differential-recurrence relations for $\langle Y_{lm}\rangle(t)$ becomes

$$\tau_N \frac{d}{dt}\langle Y_{lm}\rangle(t) = \sum_{r=-2}^{2}\sum_{s=-2}^{2} e_{lm;l+r\,m+s}(t)\langle Y_{l+r\,m+s}\rangle(t), \qquad (6)$$

where the coefficients $e_{lm;l+r\,m+s}(t)$ are now time-dependent and are given explicitly in Appendix A, $\tau_N = \tau_0 \sigma(\alpha + \alpha^{-1})$ is the free rotational diffusion time of the magnetization, and $\tau_0 = (2\gamma\mu_0 M_S D_\parallel)$ is a normalizing time. By using Eq. (4) and the definition of the spherical harmonics of first rank, viz., [42]

$$Y_{10}(\vartheta,\varphi) = \sqrt{\frac{3}{4\pi}} \cos\vartheta,$$

$$Y_{1\pm1}(\vartheta,\varphi) = \mp\sqrt{\frac{3}{8\pi}} \sin\vartheta e^{\pm i\varphi},$$

the magnetization $M_H(t) = M_S \langle \cos\Theta \rangle(t)$ in the direction of the ac driving field $\mathbf{H}\cos\omega t$ may be formally expressed via the statistical moments $\langle Y_{10} \rangle(t)$ and $\langle Y_{11} \rangle(t)$ as

$$M_H(t) = M_S \sqrt{\frac{4\pi}{3}} \left\{ \gamma_3 \langle Y_{10} \rangle(t) - \sqrt{2}\,\mathrm{Re}\left[(\gamma_1 - i\gamma_2)\langle Y_{11} \rangle(t)\right] \right\}. \tag{7}$$

However, due to the sinusoidal term in the applied field $\mathbf{H}_0 + \mathbf{H}\cos\omega t$, the stationary response of $M_H(t)$ must, in general, be developed in a Fourier series because with the notable exception of the linear response all harmonics of the ac field will now be involved, viz.,

$$M_H(t) = M_S \sum_{k=-\infty}^{\infty} m_1^k(\omega) e^{ik\omega t}, \tag{8}$$

where the Fourier coefficients $m_1^k(\omega)$ of the $k$th harmonic of $M_H(t)$ are given by

$$m_1^k(\omega) = \sqrt{\frac{2\pi}{3}} \left[ \sqrt{2}\gamma_3 c_{10}^k(\omega) + (\gamma_1 + i\gamma_2) c_{1-1}^k(\omega) - (\gamma_1 - i\gamma_2) c_{11}^k(\omega) \right] \tag{9}$$

and $c_{lm}^k(\omega)$ are themselves the Fourier coefficients in a Fourier series development in the time of the average spherical harmonics

$$\langle Y_{nm} \rangle(t) = \sum_{k=-\infty}^{+\infty} c_{nm}^k(\omega) e^{ik\omega t}. \tag{10}$$

The coefficients $c_{lm}^k(\omega)$ can then be evaluated using matrix continued fractions as described in Appendix B. Equation (8) includes the linear response as a special case, $\xi \to 0$, whereupon all higher harmonics may be discarded in Eq. (8) and only the term $m_1^1(\omega)$ linear in $\xi$ remains.

Having determined the Fourier amplitudes $m_1^k(\omega)$, we have $M_H(t)$ and other related parameters such as the dynamic susceptibilities, etc. This procedure will also yield the DMH loop representing a parametric plot of the stationary time-dependent magnetization as a function of the ac field, i.e., $M_H(t)$ vs. $H(t) = H\cos\omega t$, and the area enclosed by the loop, viz.,

$$A = -v\mu_0 \oint M_H(t)\,dH(t). \tag{11}$$

Equation (11) represents the energy loss per nanomagnet in one cycle of the ac field. The physical meaning of $A$ is that it determines the so-called specific absorption rate (SAR) defined as $SAR = \omega A/(2\pi)$. Here we shall calculate (because of its direct relation to the complex susceptibility) the *normalized* area of the DMH loop $A_n = A/(4v\mu_0 M_S H)$ given by [43]

$$A_n = -\frac{1}{4M_S H} \oint M_H(t) dH(t) = -\frac{\pi}{2} \text{Im}(m_1^1). \qquad (12)$$

The DMH phenomenon (originally predicted in nanomagnets by Ignachenko and Gekht [44]) is of much practical interest since it occurs in magnetic information storage and magnetodynamic hyperthermia occasioned by induction heating of nanomagnets.

The vectors $\mathbf{H}_0$, $\mathbf{H}$, and $\mathbf{e}_P$ (as defined in spherical polar coordinates in Fig. 1) are assumed throughout to lie in the equatorial or $XY$ plane with colatitudes $\vartheta_\xi = \pi/2$ and $\vartheta_P = \pi/2$, respectively, so that the orientations of $\mathbf{H}_0$, $\mathbf{H}$, and $\mathbf{e}_P$ are entirely specified by the azimuthal angles of the applied fields $\varphi_\xi$ and spin polarization $\varphi_P$, respectively. The values $\varphi_\xi = \varphi_P = 0$ correspond to the particular configuration whereby the vectors $\mathbf{H}_0$, $\mathbf{H}$, and $\mathbf{e}_P$ are all directed along the easy ($X$-)axis. The spin polarization azimuthal angle $\varphi_P$, biaxiallity parameter $\delta$, spin-polarization factor $P$, and damping $\alpha$ selected are $\varphi_P = 0$, $\delta = 20$, $P = 0.3$ ($P \approx 0.3 \div 0.4$ are typical values for ferromagnetic metals [20]), and $\alpha = 0.01$ (for high damping $\alpha \geq 1$, the STT effects become very small [28]). For $D_{\parallel} = 0.034$, $\gamma = 2.2 \times 10^5 \text{mA}^{-1}\text{s}^{-1}$, $M_S \approx 1.4 \times 10^6 \text{Am}^{-1}$ (cobalt), we have $\tau_0 \approx 4.8 \cdot 10^{-11}$ s. Furthermore, for $v \sim 10^{-24}$ m$^3$ and $T \sim 293$ K, the dc and ac field parameters $\xi_0$ and $\xi$ are of the order of unity for $H_0, H \sim kT/(v\mu_0 M_S) \approx 2.3 \times 10^3$ Am$^{-1}$.

### III. LINEAR DYNAMIC SUSCEPTIBILITY

For a *weak* ac field, $\xi \to 0$, all nonlinear effects in the response may be ignored, so that the magnetization $M_H(t)$ is simply given by the linear response

$$M_H(t) = M_0 + \text{Re}\{\chi(\omega)\xi e^{i\omega t}\}, \qquad (13)$$

where

$$M_0 = M_S \langle \cos\Theta \rangle_{st} = M_S m_1^0(\omega) = \frac{\omega}{2\pi} \int_0^{2\pi/\omega} M_H(t) dt$$

is the stationary time- and frequency-independent magnetization, $\langle \ \rangle_{st}$ is the statistical average, and $\chi(\omega) = 2m_1^1(\omega)/\xi$ is the linear dynamic susceptibility which is independent of the ac field strength. The corresponding plots of the real and imaginary parts of the normalized linear susceptibility $\chi(\omega)/\chi$ vs. $\omega\tau_N$ are shown in Fig. 2, where $\chi = \chi(0)$ is the static susceptibility. Just as with the zero STT case, analysis and subsequent interpretation of the linear response radically simplifies at low frequencies because the low-frequency behavior of $\chi(\omega) = \chi'(\omega) - i\chi''(\omega)$ can then be accurately described by a *single* Lorentzian, viz.,

$$\frac{\chi(\omega)}{\chi} \approx \frac{1-\Delta}{1+i\omega\tau} + \Delta. \qquad (14)$$

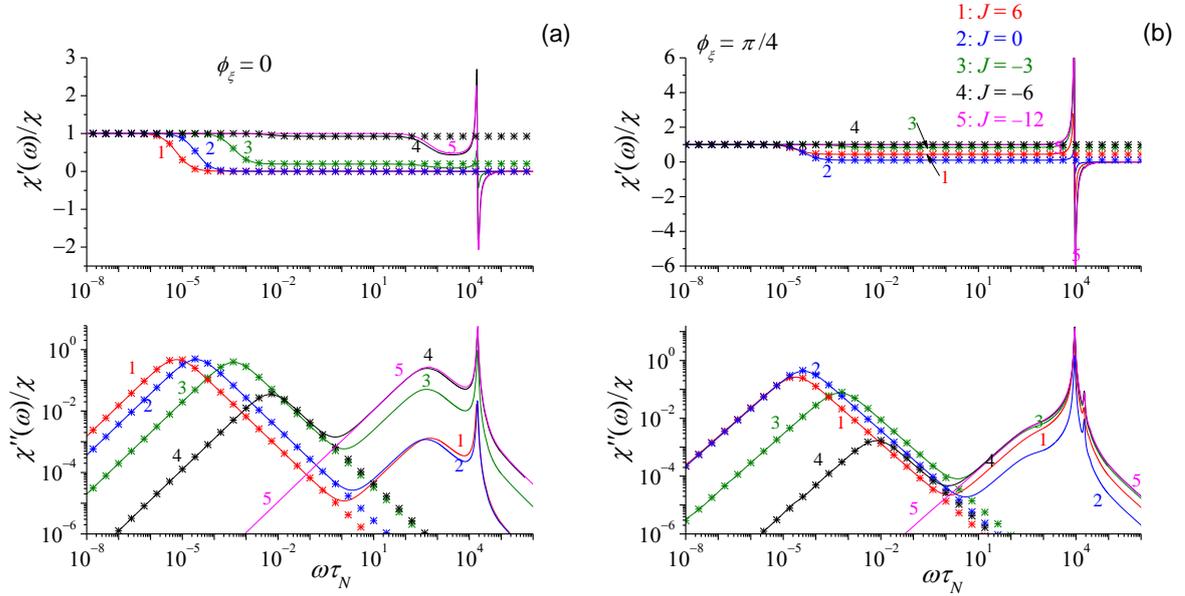

FIG. 2. (Color on line) Real and imaginary parts of the normalized linear susceptibility $\chi(\omega)/\chi$ vs. the normalized frequency $\omega\tau_N$ for various spin-polarized current parameters $J = 6, 0, -3, -6, -12$ and for various orientations of the applied fields $\varphi_\xi = 0$ (a) and $\varphi_\xi = \pi/4$ (b) with the anisotropy parameter $\sigma = 20$ and the dc field parameter $\xi_0 = 2$. Solid lines: matrix continued fraction solution. Asterisks: Approximate Eq. (14) with the reversal time $\tau = \lambda_1^{-1}$ calculated using the independent method of Ref. [28].

In Eq. (14), $\tau$ is the longest (overbarrier) relaxation time without the ac external field, and $\Delta$ is a parameter accounting for the mid- and high-frequency relaxation processes. Now $\tau$ is related to the frequency $\omega_{\max}$ of the low-frequency peak in the loss spectrum $-\text{Im}[\chi(\omega)]$, where it attains a maximum, and the half-width $\Delta\omega$ of the spectrum of the real part of the susceptibility $\text{Re}[\chi(\omega)]$ via

$$\tau \approx \omega_{\max}^{-1} \approx \Delta\omega^{-1}. \qquad (15)$$

Since $\tau$ is the magnetization reversal time (effectively the inverse escape rate), it can be associated with the inverse of the smallest nonvanishing eigenvalue $\lambda_1$ of the Fokker-Planck operator as comprehensively described in Ref. [28]. Comparison of $\tau$ as extracted from the spectra $\chi(\omega)$ via Eq. (14) with $\tau = \lambda_1^{-1}$ calculated independently via $\lambda_1$ of the Fokker-Planck operator [28] shows that both methods yield identical results. Also varying of the material and geometrical model parameters may alter the reversal time $\tau$ by orders of magnitude concurring with experimental results [11]. The dependence of $\tau$ on the model parameters (damping $\alpha$, spin-polarized current parameter $J$, the external field strength and orientation, etc.) has been given in Ref. 28.

The main features of the normalized linear susceptibility plots are as follows. For $J$ of intermediate magnitudes, the overall picture is more or less similar to that for $J = 0$, i.e., we have the usual low-frequency overbarrier (interwell) relaxation, mid-frequency intrawell relaxation, and high-frequency ferromagnetic resonance (FMR) behavior in a biaxial potential. Thus, we have, in general, three dispersion regions in $\text{Re}[\chi(\omega)/\chi]$ and three corresponding absorption bands in the magnetic loss spectrum $-\text{Im}[\chi(\omega)/\chi]$ (see Fig. 2). The broad low-frequency peak in $-\text{Im}[\chi(\omega)/\chi]$ corresponds to *slow* reversal of the magnetization vector over the potential barriers and is accurately described by the approximate Eq. (14). The most pronounced STT effect is that the decrease of *J* from large positive values *initially* shifts the low-frequency relaxation peak to *lower* frequencies until the peak frequency $\omega_{\max}$ reaches a minimum at some intermediate value of *J* above which the peak is shifted to *higher* frequencies (only the shift to higher frequencies is shown in Fig. 2). This minimum frequency peak corresponds to the particular situation, where the STT has annulled the effect of the dc bias field so that the effective potential has equal well depths. This corresponds to the maximum relaxation time at a definite value of $J_{\max}$, which has been depicted graphically in Fig. 10 of Ref. [28]. For high positive or negative *J*, the magnitude of the low-frequency peak in $-\text{Im}[\chi(\omega)/\chi]$ decreases until it merges with the mid-frequency peak, signifying that the overbarrier relaxation process has been completely extinguished due to the action of STT. Thus, *high magnitude spin-polarized current seems to have virtually the same effect on the magnetization reversal as that of a strong dc bias field in single domain ferromagnetic particles at zero STT* [45,46] (see also [27], Chap. 9). Here at a certain critical value of that field [45,46] which is much less than the nucleation field the integral relaxation time (area under the curve of the magnetization decay) diverges exponentially from the overbarrier relaxation time due to the depletion of the population of the shallowest well of the potential by the dc field. This event is also signified by the virtual disappearance of the low-frequency peak in the magnetic loss spectrum. Thus, all that remains are the dynamical processes within the wells. The explanation appears to be that the high positive or negative *J* reduces the effective potential barrier so that the overbarrier time decreases. Regarding the second peak at intermediate frequencies, this is due to fast near-degenerate exponential decays in the wells of the effective potential and comprises the usual longitudinal *intrawell* relaxation. Lastly, we see a third FMR peak at the Larmor frequency $\omega_{pr}$. The origin of this peak lies in the magnetization precession in an effective field due to both the anisotropy and applied dc field. Notice that the susceptibility is strongly influenced by the azimuthal angle $\varphi_\xi$ of the applied ac field. For $\varphi_\xi \neq 0$, high-frequency resonant harmonic modes are discernible in the FMR band generating a comb-like structure with characteristic frequencies $n\omega_{pr}, n = 2,3,...$ reminiscent of that which occurs in inertia-corrected dielectric relaxation of polar molecules at THz frequencies under the influence of a mean field potential [27,47]. This comb-like structure virtually disappears, however, for $\varphi_\xi = 0$. The STT has no effect on the mid-frequency and FMR regions with any apparent changes being purely an artifact of the normalization.

## IV. NONLINEAR RESPONSE

In strong applied ac fields, $\xi > 1$, pronounced frequency-dependent nonlinear effects occur (see Figs. 3 and 4 illustrating the dependence of the nonlinear response on the ac field strength parameter $\xi$). In contrast to the linear response, the stationary *time-independent* but now *frequency-dependent* magnetization $M_0(\omega) = M_S m_1^0(\omega)$ and the nonlinear dynamic susceptibility $\chi(\omega) = 2m_1^1(\omega)/\xi$ as well as all other higher harmonics $m_1^k(\omega)$ with $k > 1$ now strongly depend on the magnitude $\xi$ of the ac field. Moreover, for given $\xi$, all $m_1^k(\omega)$ also markedly depend on the azimuthal angle $\varphi_\xi$, dc bias field $\xi_0$, anisotropy parameters $\sigma$ and $\delta$, damping $\alpha$, and spin-polarized current parameter $J$, while the time-independent component of the magnetization $M_0$ alters profoundly leading to new nonlinear effects. In particular, that component in typical nonlinear fashion becomes *dependent on both the amplitude and frequency of the ac field* (see Fig. 3). Such behavior is in sharp contrast to that of nanomagnets in an ac field omitting STT, where one must also apply a dc bias in combination with a strong ac field in order to observe the frequency dependence of the dc response. This effect being due to *entanglement* of the nonlinear ac and dc responses [48]. However, with STT included the ac field amplitude and frequency dependence of $M_0$ always exists even for zero dc bias field, i.e., $\xi_0 = 0$. Hence, the spin-polarized current seems to have the same effect on $M_0(\omega)$ as that of a dc bias field at zero STT. Furthermore the dc response in Fig. 3 is not an odd function of $J$ due to the nature of the non-conservative potential $\Phi$.

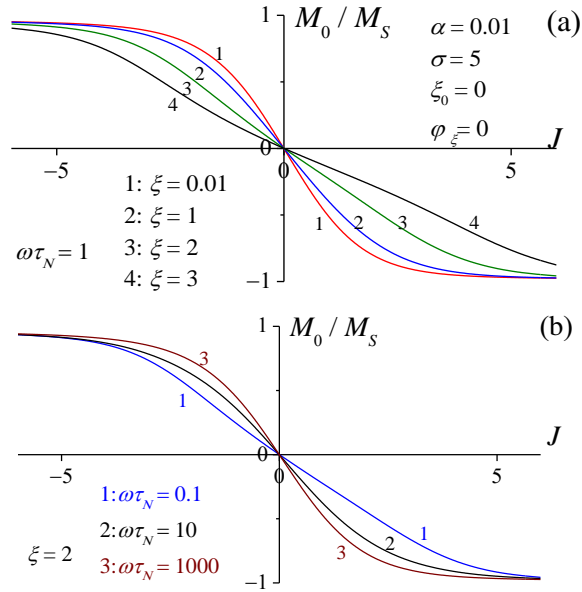

FIG. 3. (Color on line) (a) Time-independent (dc) component of the magnetization $M_0/M_S$ vs. the spin-polarized current parameter $J$ for (a) various ac field amplitudes $\xi = 0.01, 1, 2,$ and $3$ ($\xi = 0.01$ represents linear response) and $\omega\tau_N = 1$ and for (b) various frequencies $\omega\tau_N = 10^{-1}, 10,$ and $10^3$ at $\xi = 2$ at $\sigma = 5$, $\varphi_\xi = 0$, and $\xi_0 = 0$. Solid lines: the matrix continued fraction solution.

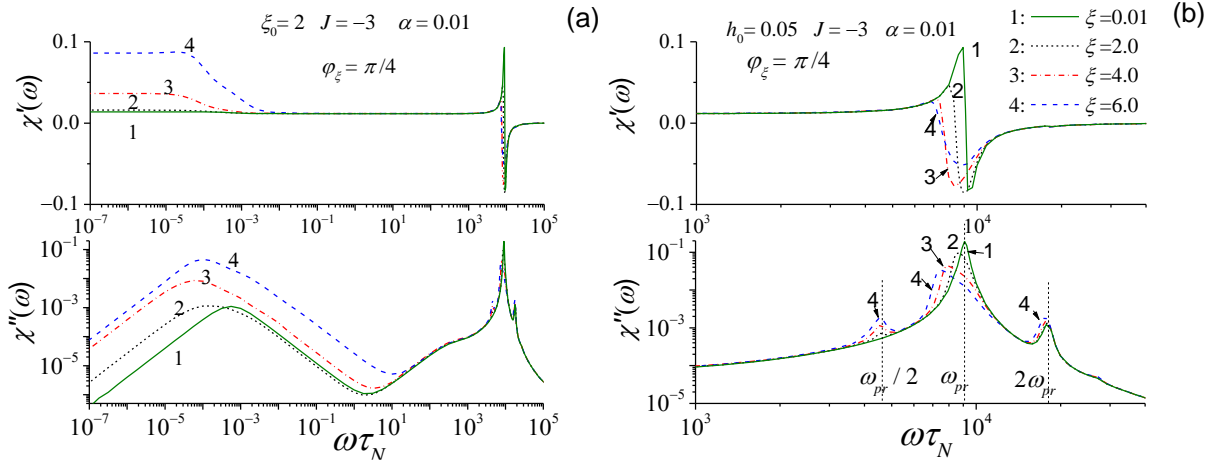

FIG. 4. (Color on line) (a) Real and imaginary parts of the nonlinear susceptibility $\text{Re}[\chi(\omega)]$ and $-\text{Im}[\chi(\omega)]$ vs. $\omega\tau_N$ for various ac field amplitudes $\xi$ and $J = -3$, $\varphi_\xi = \pi/4$ $\sigma = 20$, and $\xi_0 = 2$. Solid and dashed lines: linear and nonlinear response, respectively, using the matrix continued fraction solution. (b) The high-frequency parts of the spectra alone.

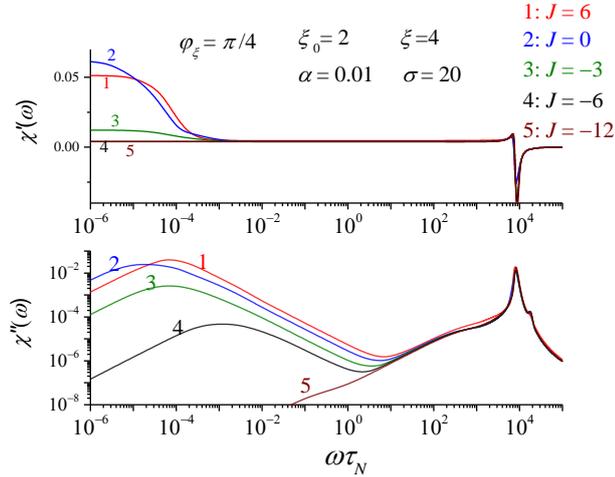

FIG. 5. (Color on line) Real and imaginary parts of the nonlinear susceptibility $\text{Re}[\chi(\omega)]$ and $-\text{Im}[\chi(\omega)]$ vs $\omega\tau_N$ for various spin-polarized current parameter $J = 6, 0, -3, -6, -12$ and $\sigma = 20$, $\varphi_\xi = \pi/4$, $\xi_0 = 2$, and $\xi = 4$. Solid lines: the matrix continued fraction solution.

Now in strong ac fields, it appears that the low-frequency band of $-\text{Im}[\chi(\omega)]$ deviates substantially from the Lorentzian shape so that it can no longer be approximated by the single Lorentzian Eq. (14). Nevertheless, Eq. (15) may still be used in order to estimate an effective magnetization reversal time $\tau$ as $\tau \approx \Delta\omega^{-1}$. Furthermore, as the ac field strength $\xi$ increases, the magnitude of the low-frequency peak in $-\text{Im}[\chi(\omega)]$ is enhanced (Fig. 4) and also with increasing $\xi$ the overbarrier peak on initially shifting to lower frequencies, attains a minimum frequency, thereafter shifting to higher frequencies. Omitting STT, this minimum frequency peak will occur at $\xi \sim \xi_0$. However, this is not true when STT is included as the STT acts in combination with the applied field (Fig. 4). We remark that the reversal time $\tau$ may also be evaluated from the spectra of both the dc component $M_0(\omega)$ and the higher-order harmonics $m_1^k(\omega)$ with $k > 1$ because the low-frequency parts of these spectra are themselves dominated by overbarrier relaxation processes with the characteristic time $\tau$. Now as seen in Fig. 4 (b), again with increasing $\xi$, the magnitude of the main FMR peak at the precession frequency $\omega_{pr}$ decreases and also broadens showing pronounced saturation effects. Moreover, a new high-frequency dispersion of resonant character near the frequency $\sim \omega_{pr}/2$ due to *parametric resonance* appears just as that commonly occurring in nonlinear oscillators driven by an ac external force. Nevertheless, the high-frequency ($\omega \gg \omega_{pr}$) behavior of the spectrum remains virtually unchanged (see Fig. 4). Parametric excitations of a current-biased nanomagnet by a microwave magnetic field were observed recently by Urazhdin *et al.* [34] amply demonstrating that this phenomenon can be used to determine dynamical properties of nanomagnets.

The nonlinear susceptibility for various $J$ (Fig. 5) exhibits many of the same characteristics as the corresponding linear susceptibility (Fig. 2), i.e., three dispersion regions in $\text{Re}[\chi(\omega)/\chi]$ and three corresponding absorption bands in the magnetic loss spectrum $-\text{Im}[\chi(\omega)/\chi]$ for $J$ of intermediate magnitude, merging of the overbarrier peak with the mid-frequency peak for high positive or negative $J$, and virtually no STT effect in the mid-frequency and FMR regions. For $J < 0$ the magnitude of the low-frequency overbarrier peak decreases and the peak is shifted to higher frequencies, while for $J > 0$ the magnitude of this peak increases and the peak also shifts to higher frequencies.

## V. DYNAMIC MAGNETIC HYSTERESIS

For a *weak* ac field, $\xi \to 0$, the DMH loops $[m(t) = M_H(t)/M_S$ vs. reduced ac field $h(t) = \cos \omega t$ ] are ellipses with normalized area $A_n$ given by Eq. (12); the behavior of $A_n \sim \chi'' \sim -\text{Im}(m_1^1)$ being similar [cf. Eq. (12)] to that of the magnetic loss $\chi''(\omega)$ (see Figs. 2,4, and 5). Now for *moderate* ac fields, $\xi \approx 1$, the DMH loops still have an ellipsoidal shape implying that only a few harmonics actually contribute to the nonlinear response. However, in *strong* ac fields, $\xi > 1$, the loop shape alters substantially (see Figs. 6-9). In Fig. 6, the DMH loops are plotted for various values of $J$ and ac amplitude $\xi$ showing that both their shapes and their areas alter as these parameters vary. The pronounced frequency dependence of the DMH is highlighted in Figs. 8 and 9 for $\varphi_\xi = 0$ and $\varphi_\xi = \pi/4$, respectively, which also illustrates their azimuthal angle dependence. In the low frequency band (Figs. 8a-8c and 9a-9c), the negative and positive $J$ shifts the DMH loops to the left and right respectively. Moreover, at low frequencies, the field changes are *quasi-* *adiabatic*, so that the magnetization reverses due to the *cooperative* shuttling action of thermal agitation, STT, and ac field. In contrast at high frequencies (see Figs. 8e, 8f, 9e and 9f), the origin of the DMH lies in the *resonant* dispersion and absorption in the FMR band. Here, the phase difference $\Delta\phi$ between $M_H(t)$ and $H(t)$, governing *loop orientation*, may undergo a pronounced variation in the very high frequency FMR band as is typical of a resonant process. In particular, the phase difference may exceed $\pi/2$ (see, e.g., [49]). Obviously, this large resonant effect does not exist at low and intermediate frequencies, where $\Delta\phi$ is always less than $\pi/2$. At FMR frequencies, DMH occurs due to the *resonant* behavior of the nonlinear response (see Fig. 8f) and under such conditions the switching may be termed "resonant", leading naturally to the concept of *resonant switching of the magnetization*. Since the resonant DMH occurs at very high (GHz) frequencies, the magnetization switching is, therefore, for the most part governed by the frequency of the external driving field, or equivalently, the rate of change of the amplitude of the latter. Hence, the magnetization may be advantageously switched in this situation, because the field needed to reverse it is then much smaller than the quasi-static coercive force [49].

By plotting the normalized area $A_n$ vs. the spin-polarized current (Fig. 10), $A_n$ can invariably be represented as a bell curve with the height, width, and center of the peak determined by the various parameters. This is similar to a plot of $A_n$ vs. the dc bias field strength $\xi_0$ except that the latter will always have the center of the peak along $\xi_0 = 0$. In Fig. 10a, on increasing the ac field strength $\xi$, $A_n$ also increases and the range of *J*, for which a significant DMH loop area exists, broadens. In *strong* ac fields, $\xi > 1$, the normalized area alters substantially (see Fig. 10a). Nevertheless, $A_n$ is still determined by $-\text{Im}(m_1^1)$ [cf. Eq. (12)]. Thus $A_n$ strongly depends on the frequency $\omega$, the angles $\varphi_\xi$ and $\varphi_P$, ac and dc bias field amplitudes $\xi$ and $\xi_0$ as well as the anisotropy parameters $\sigma$ and $\delta$, damping $\alpha$, and the spin-polarized current parameter *J*. In Fig. 10b, on increasing the driving frequency, the normalized area initially increases, reaches a maximum, and then decreases.

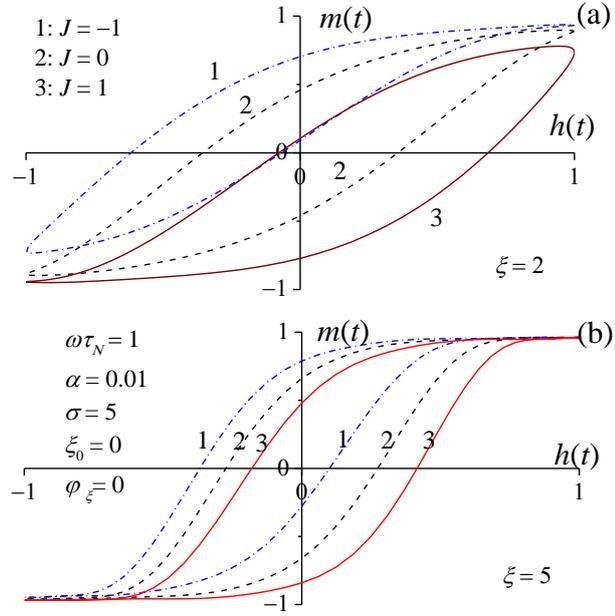

FIG. 6. (Color on line) DMH loops for various spin-polarized current parameter $J = -1, 0, 1$ and (a) $\xi = 2$ and (b) $\xi = 5$ with $\omega\tau_N = 1$, $\varphi_\xi = 0$, $\alpha = 0.01$, $\sigma = 5$, and $\xi_0 = 0$ (calculated using the matrix continued fraction solution).

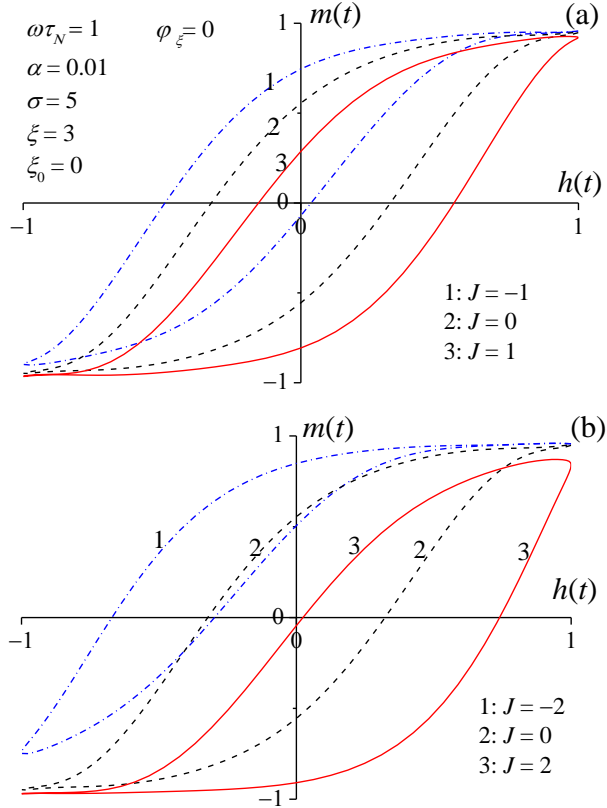

FIG. 7. (Color on line) DMH loops for various spin polarized current parameter $J = -1, 0, 1$ (a) and $J = -2, 0, 2$ (b) with $\omega\tau_N = 1$, $\varphi_\xi = 0$, $\alpha = 0.01$, $\xi_0 = 0$, $\xi = 3$, and $\sigma = 5$.

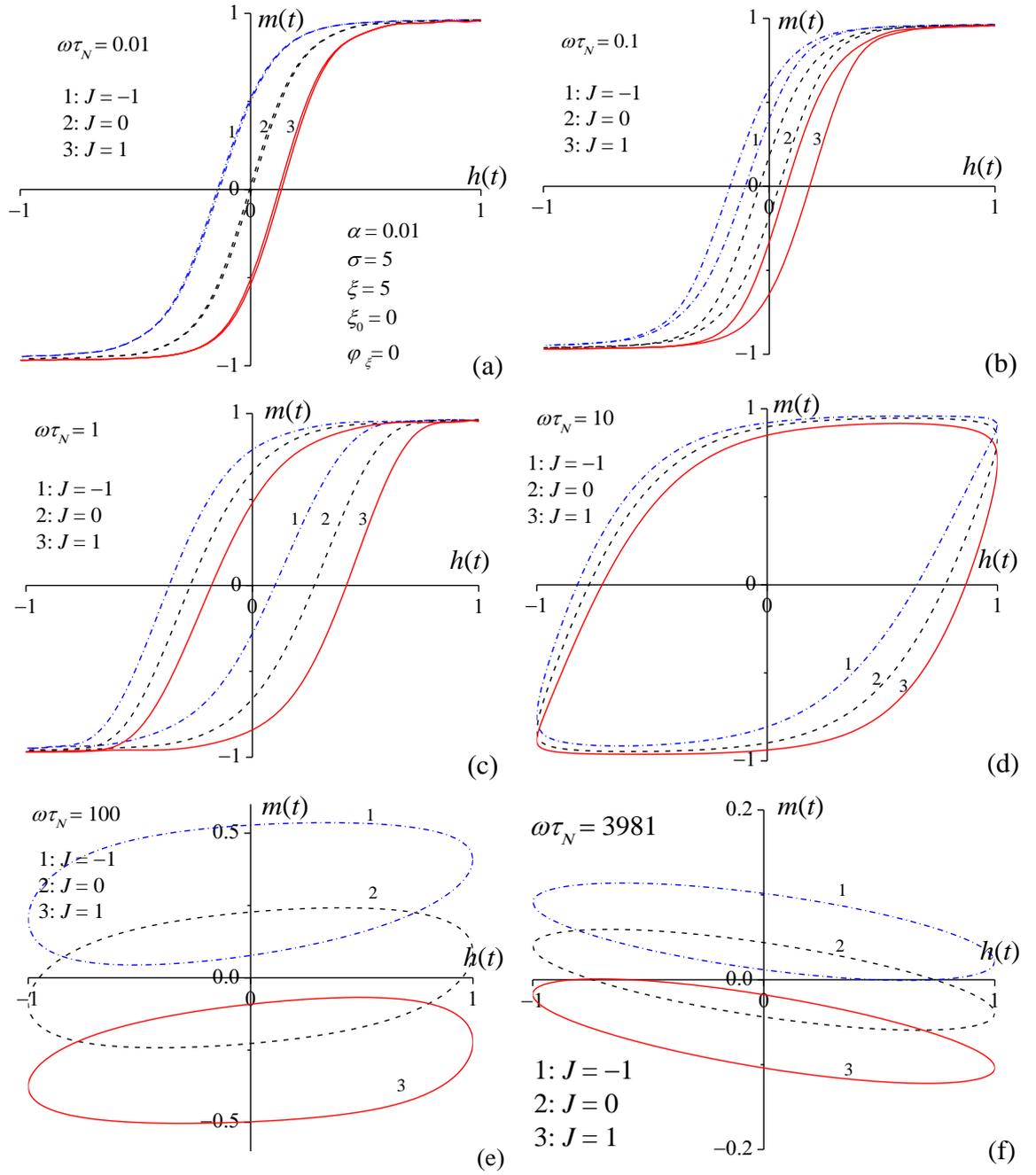

FIG. 8. (Color on line) DMH loops for various spin polarized current parameters $J = -1, 0, 1$ and frequencies $\omega\tau_N = 10^{-2}$ (a), $10^{-1}$ (b), 1 (c), 10 (d), $10^2$ (e), and 3981 (f) with $\varphi_\xi = 0$, $\alpha = 0.01$, $\xi_0 = 0$, $\xi = 5$, and $\sigma = 5$.

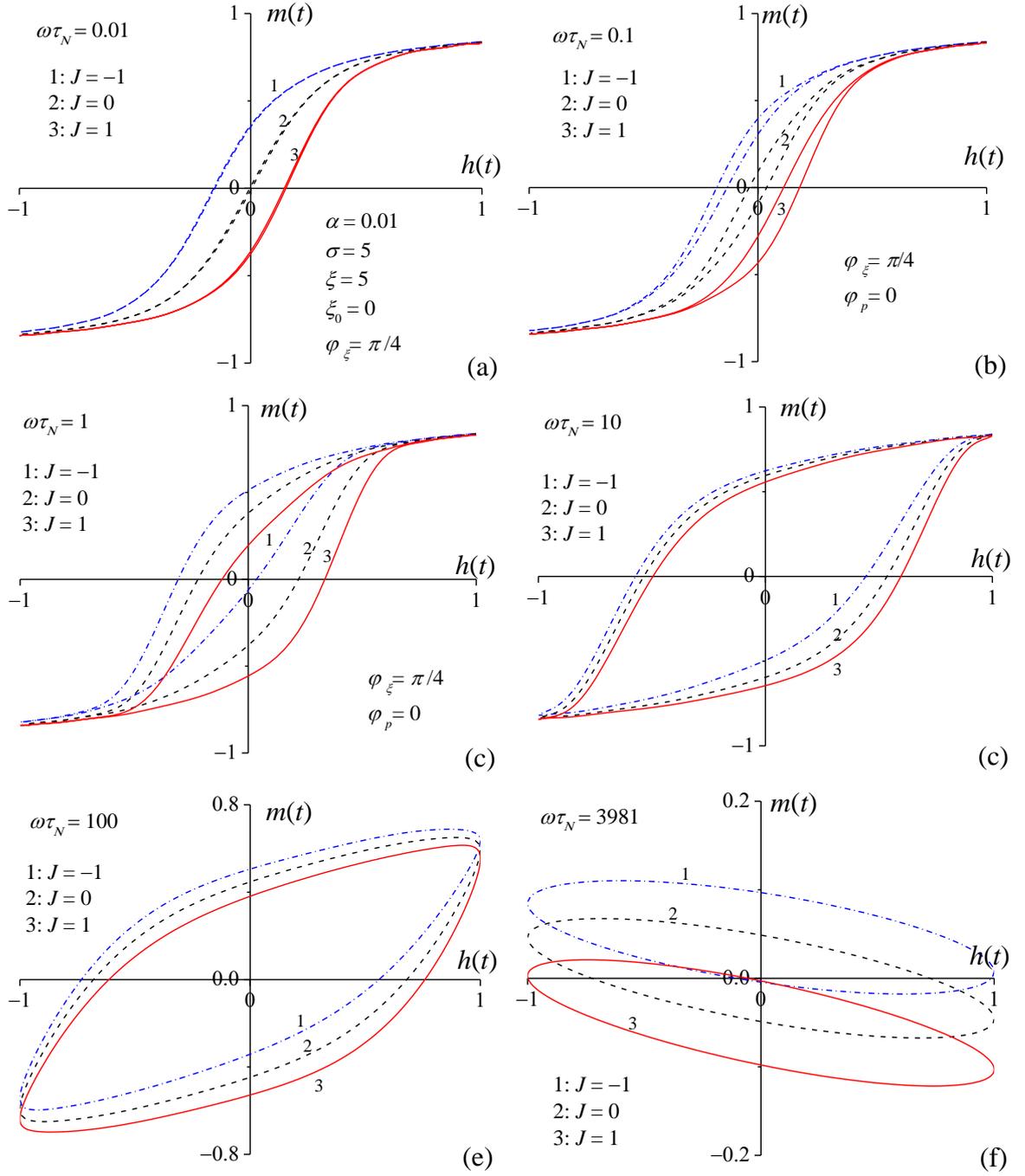

FIG. 9. (Color on line) DMH loops for various spin-polarized current parameters $J = -1, 0, 1$ and frequencies $\omega\tau_N = 10^{-2}$ (a), $10^{-1}$ (b), 1 (c), 10 (d), $10^2$ (e), and 3981 (f) with $\varphi_\xi = \pi/4$, $\alpha = 0.01$, $\xi_0 = 0$, $\xi = 5$, and $\sigma = 5$ (calculated using the matrix continued fraction solution).

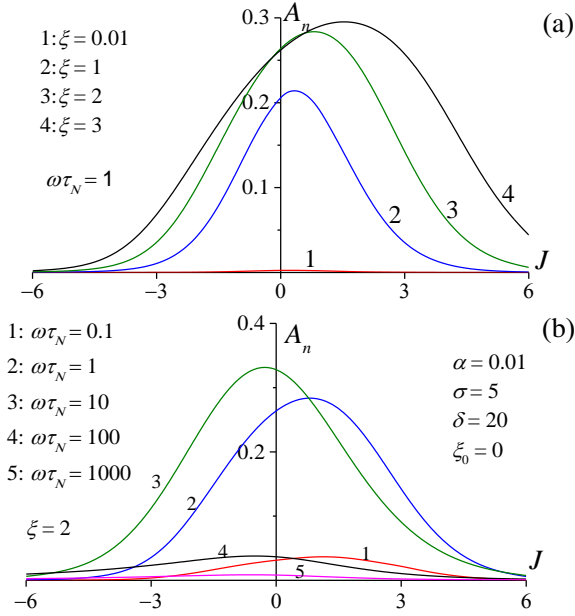

FIG. 10. (Color on line) *Normalized* area of the DMH loop $A_n$, Eq. (12), vs. spin-polarized current parameter *J* (a) for various ac field amplitudes $\xi$ and $\omega\tau_N = 1$ and (b) for various frequencies $\omega\tau_N$ and $\xi = 2$ with $\varphi_\xi = 0$, $\alpha = 0.01$, $\xi_0 = 0$, and $\sigma = 5$ (calculated using the matrix continued fraction solution).

## VI. CONCLUSIONS

We have treated STT effects on the ac stationary forced response of nanoscale ferromagnets driven by an ac magnetic field of arbitrary strength using a nonperturbative approach originally developed [27] for nanomagnets omitting STT. Our method, based on the solution of the differential-recurrence relation for the infinite hierarchy of statistical moments generated by either the Langevin or Fokker-Planck equations as augmented by STT terms, indicates that STT profoundly alters the nonlinear response of a nanomagnet leading to new effects. Furthermore, the statistical moment approach holds for the *most comprehensive* formulation of the generic nanopillar model (Fig. 1), i.e., for *arbitrary directions* of the dc and ac external fields allowing us to treat STT effects on frequency-dependent characteristics under conditions which are otherwise inaccessible. Clearly, at low damping, the stationary response to an ac driving field is very sensitive to both the intensity of the spin-polarized current and the frequency and amplitude of that field owing to the *intrinsic* coupling between the magnetization precession and its thermally activated reversal. Furthermore, our calculations, since they are valid for ac fields of *arbitrary* strength and orientation, quantify the role played by STT in nonlinear phenomena in nanoscale ferromagnets such as nonlinear stochastic resonance and dynamic magnetic hysteresis, nonlinear ac field effects on the switching field curves, etc., where perturbation theory is no longer valid. In addition, the moment method yields the response for *all frequencies of interest* including very high frequencies covering the ferromagnetic resonance (GHz) range exemplifying various nonlinear phenomena such as parametric resonance and higher harmonic generation (which we hope will stimulate new experiments). Hence, the high-frequency linear and nonlinear FMR spectra (see Figs. 2 and 4) may be suitable for the purpose of explaining the line shape of STT nano-oscillators driven by ac external magnetic fields and currents. Likewise, the DMH loops and their area (yielding the Joule heating during the switching process) as well as the calculations of the effective magnetization reversal time via the low-frequency band of the magnetic loss spectra may be useful for the prediction, modeling, and interpretation of switching processes in recording techniques. Furthermore, the DMH arising from a *high-frequency* periodic signal may be exactly evaluated permitting quantitative analysis of *ultrafast switching* of the magnetization. In particular, accurate solutions in the manner outlined of the hierarchy of the statistical moment equations for a generic model are essential for the future development of both escape rate theory and stochastic dynamics simulations of the magnetization reversal process in STT systems just as they were in single domain particles. For the limit of zero STT, our results concur with established solutions for nanomagnets with biaxial anisotropy [41] while, for nonzero STT, they constitute rigorous benchmark solutions with which calculations of nonlinear response characteristics via any other approach must comply. Finally, the statistical moment method may be similarly generalized to the forced response of a nanoscale ferromagnet driven by an alternating current.

## ACKNOWLEDGMENTS

We would like to thank FP7-PEOPLE-Marie Curie Actions - International Research Staff Exchange Scheme (Project No. 295196 DMH) for financial support. Moreover, W. T. Coffey thanks Ambassade de France in Ireland for research visits to Perpignan. D. Byrne, acknowledges the SimSci Structured Ph.D. Program at University College Dublin for financial support. This research is supported by the Programme for Research In Third Level Institutions (PRTLI) Cycle 5 and co-funded by the European Regional Development Fund. We also thank P. M. Déjardin for helpful conversations.

## APPENDIX A : EXPLICIT FORM OF THE COEFFICIENTS $e_{nm;n'm'}(t)$

By applying the general approach [27,28,40] for the derivation of differential-recurrence relations from the magnetic Langevin equation (1) as specialized to the potentials Eqs. (3) and (5), we have the 25 term differential-recurrence Eq. (6) for the statistical moments $c_{lm}(t) = \langle Y_{lm} \rangle(t)$, viz.,

$$\tau_N \frac{d}{dt} c_{nm}(t) = v_{nm}^{--} c_{n-2m-2}(t) + v_{nm}^{-} c_{n-2m-1}(t) + v_{nm} c_{n-2m}(t) + v_{nm}^{+} c_{n-2m+1}(t) + v_{nm}^{++} c_{n-2m+2}(t)$$
$$+ w_{nm}^{--} c_{n-1m-2}(t) + w_{nm}^{-}(t) c_{n-1m-1}(t) + w_{nm}(t) c_{n-1m}(t) + w_{nm}^{+}(t) c_{n-1m+1}(t) + w_{nm}^{++}(t) c_{n-1m+2}(t)$$
$$+ x_{nm}^{--} c_{nm-2}(t) + x_{nm}^{-}(t) c_{nm-1}(t) + x_{nm}(t) c_{nm}(t) + x_{nm}^{+}(t) c_{nm+1}(t) + x_{nm}^{++}(t) c_{nm+2}(t) \qquad (16)$$
$$+ y_{nm}^{--} c_{n+1m-2} + y_{nm}^{-}(t) c_{n+1m-1}(t) + y_{nm}(t) c_{n+1m}(t) + y_{nm}^{+}(t) c_{n+1m+1}(t) + y_{nm}^{++} c_{n+1m+2}(t)$$
$$+ z_{nm}^{--} c_{n+2m-2}(t) + z_{nm}^{-} c_{n+2m-1}(t) + z_{nm} c_{n+2m}(t) + z_{nm}^{+} c_{n+2m+1}(t) + z_{nm}^{++} c_{n+2m+2}(t).$$

Here the coefficients $x_{nm}(t)$, $y_{nm}(t)$, etc. corresponding to the matrix elements $e_{nm;n'm'}(t)$ in Eq. (6) have the same form as those in Eq. (C1) of Ref. [28] save that they are now *time-dependent*. To keep in step with the notation of Ref. [28], we define reduced fields

$$h_0 = H_0 / (2M_S D_\|) \text{ and } h = H / (2M_S D_\|) \tag{17}$$

Thus in our particular case of a single ac forcing term, we have

$$h(t) \to h_0 + 2h\cos\omega t \to h_0 + h(e^{i\omega t} + e^{-i\omega t}) \tag{18}$$

Equation (18) then implies that for a constant field superimposed on a periodic ac one, the time-dependent coefficients $x_{nm}(t)$, etc. in Eq. (16) may be written as the sum of a dc term and one oscillating at the fundamental frequency, viz.,

$$w_{nm}(t) = w_{nm}^0 + w_{nm}^1 (e^{i\omega t} + e^{-i\omega t}),$$

$$x_{nm}(t) = x_{nm}^0 + x_{nm}^1 (e^{i\omega t} + e^{-i\omega t}),$$

$$y_{nm}(t) = y_{nm}^0 + y_{nm}^1 (e^{i\omega t} + e^{-i\omega t}),$$

etc. The various coefficients are then given by

$$x_{nm}^0 = -\frac{n(n+1)}{2} + i\frac{m\sigma h_0 \gamma_3}{\alpha} + i\sqrt{\frac{\pi}{3}} m b_P J Y_{10}^*(\vartheta_P, \varphi_P)$$
$$+ \frac{n(n+1)-3m^2}{(2n-1)(2n+3)}\left[-\sigma\left(\frac{1}{2}+\delta\right) + \frac{2\pi c_P b_P J}{3\alpha}\left(Y_{10}^{*2}(\vartheta_P, \varphi_P) + Y_{11}^*(\vartheta_P, \varphi_P)Y_{1-1}^*(\vartheta_P, \varphi_P)\right)\right],$$

$$x_{nm}^1 = i\frac{m\sigma h \gamma_3}{\alpha},$$

$$x_{nm}^{\pm 0} = \sqrt{(1 \pm n \pm m)(n \mp m)}\left\{(\gamma_1 \mp i\gamma_2)\frac{i\sigma h_0}{2\alpha}\right.$$
$$\left. \mp i\sqrt{\frac{\pi}{6}} b_P J Y_{1\pm1}^*(\vartheta_P, \varphi_P) + \frac{\sqrt{2}\pi c_P b_P J(1 \pm 2m)}{3\alpha(2n-1)(2n+3)} Y_{10}^*(\vartheta_P, \varphi_P)Y_{1\pm1}^*(\vartheta_P, \varphi_P)\right\},$$

$$x_{nm}^{\pm 1} = \sqrt{(1 \pm n \pm m)(n \mp m)}(\gamma_1 \mp i\gamma_2)\frac{i\sigma h}{2\alpha},$$

$$x_{nm}^{\pm\pm 0} = -\frac{3\sqrt{(n \pm m+1)(n \pm m+2)(n \mp m-1)(n \mp m)}}{4(2n-1)(2n+3)}\left(\sigma + \frac{4\pi c_P b_P J}{3\alpha}Y_{1\pm1}^{*2}(\vartheta_P, \varphi_P)\right),$$

$$y_{nm}^0 = \sqrt{\frac{(n+1)^2 - m^2}{(2n+1)(2n+3)}}\left\{-\frac{im\sigma}{\alpha}\left(\frac{1}{2}+\delta\right) - n\sigma h_0 \gamma_3 + \frac{b_P J n}{\alpha}\sqrt{\frac{\pi}{3}}Y_{10}^*(\vartheta_P, \varphi_P)\right.$$
$$\left. - \frac{im 4\pi c_P b_P J}{3}\left(Y_{10}^{*2}(\vartheta_P, \varphi_P) + Y_{11}^*(\vartheta_P, \varphi_P)Y_{1-1}^*(\vartheta_P, \varphi_P)\right)\right\},$$

$$y_{nm}^1 = -n\sqrt{\frac{(n+1)^2 - m^2}{(2n+1)(2n+3)}}\sigma h \gamma_3$$

$$y_{nm}^{\pm 0} = \sqrt{\frac{(1+n\pm m)(2+n\pm m)}{(1+2n)(3+2n)}} \left\{ \frac{b_P J n}{\alpha} \sqrt{\frac{\pi}{6}} Y_{1\pm 1}^*(\vartheta_P, \varphi_P) \right.$$

$$\left. \pm \frac{n}{2} \sigma h_0 (\gamma_1 \mp i\gamma_2) \pm \frac{i\sqrt{2}\pi c_P b_P J(n\mp 2m)}{3} Y_{10}^*(\vartheta_P, \varphi_P) Y_{1\pm 1}^*(\vartheta_P, \varphi_P) \right\},$$

$$y_{nm}^{\pm 1} = \pm \frac{n}{2} \sqrt{\frac{(1+n\pm m)(2+n\pm m)}{(1+2n)(3+2n)}} \sigma h (\gamma_1 \mp i\gamma_2)$$

$$y_{nm}^{\pm\pm 0} = \mp i \left( \frac{\sigma}{4\alpha} - \frac{\pi c_P b_P J}{3} Y_{1\pm 1}^{*2}(\vartheta_P, \varphi_P) \right) \sqrt{\frac{(1+n\pm m)(2+n\pm m)(3+n\pm m)(n\mp m)}{(1+2n)(3+2n)}},$$

$$w_{nm}^0 = \sqrt{\frac{n^2-m^2}{4n^2-1}} \left\{ -\frac{im\sigma}{\alpha}\left(\frac{1}{2}+\delta\right) + (n+1)\sigma h_0 \gamma_3 \right.$$

$$\left. -\frac{b_P J(n+1)}{\alpha}\sqrt{\frac{\pi}{3}} Y_{10}^*(\vartheta_P, \varphi_P) - \frac{im 2\pi c_P b_P J}{3}\left(Y_{10}^{*2}(\vartheta_P, \varphi_P) + Y_{11}^*(\vartheta_P, \varphi_P)Y_{1-1}^*(\vartheta_P, \varphi_P)\right) \right\},$$

$$w_{nm}^1 = \sqrt{\frac{n^2-m^2}{4n^2-1}}(n+1)\sigma h \gamma_3$$

$$w_{nm}^{\pm 0} = \sqrt{\frac{(n\mp m)(n\mp m-1)}{4n^2-1}} \left\{ \frac{b_P J(n+1)}{\alpha} \sqrt{\frac{\pi}{6}} Y_{1\pm 1}^*(\vartheta_P, \varphi_P) \right.$$

$$\left. \pm \frac{n+1}{2}\sigma h_0(\gamma_1 \mp i\gamma_2) \pm \frac{i\sqrt{2}\pi c_P b_P J(n+1\pm 2m)}{3} Y_{10}^*(\vartheta_P, \varphi_P) Y_{1\pm 1}^*(\vartheta_P, \varphi_P) \right\},$$

$$w_{nm}^{\pm 1} = \pm \frac{n+1}{2}\sqrt{\frac{(n\mp m)(n\mp m-1)}{4n^2-1}} \sigma h (\gamma_1 \mp i\gamma_2)$$

$$w_{nm}^{\pm\pm 0} = \pm \frac{i}{4}\left(\frac{\sigma}{\alpha} - \frac{4\pi c_P b_P J}{3} Y_{1\pm 1}^{*2}(\vartheta_P, \varphi_P)\right) \sqrt{\frac{(n\mp m-2)(n\mp m-1)(1+n\pm m)(n\mp m)}{4n^2-1}},$$

$$z_{nm} = \frac{n}{2n+3}\sqrt{\frac{[(n+1)^2-m^2][(n+2)^2-m^2]}{(2n+1)(2n+5)}}$$

$$\times \left\{ \sigma\left(\frac{1}{2}+\delta\right) - \frac{2\pi^2 c_P b_P J}{3\alpha}\left(Y_{10}^{*2}(\vartheta_P, \varphi_P) + Y_{11}^*(\vartheta_P, \varphi_P)Y_{1-1}^*(\vartheta_P, \varphi_P)\right) \right\},$$

$$z_{nm}^{\pm} = -\frac{2\sqrt{2}\pi c_P b_P J}{3\alpha} Y_{10}^*(\vartheta_P, \varphi_P) Y_{1\pm 1}^*(\vartheta_P, \varphi_P) \frac{n}{2n+3} \sqrt{\frac{[(n+1)^2-m^2](n\pm m+2)(n\pm m+3)}{(2n+1)(2n+5)}},$$

$$z_{nm}^{\pm\pm} = -\left(\frac{\sigma}{4} + \frac{\pi c_P b_P J}{3\alpha} Y_{1\pm 1}^{*2}(\vartheta_P, \varphi_P)\right) \frac{n}{2n+3} \sqrt{\frac{(n+1\pm m)(2+n\pm m)(3+n\pm m)(4+n\pm m)}{(2n+1)(2n+5)}},$$

$$v_{nm} = \frac{n+1}{2n-1}\sqrt{\frac{[(n-1)^2-m^2](n^2-m^2)}{(2n+1)(2n-3)}} \left\{ -\sigma\left(\frac{1}{2}+\delta\right) \right.$$

$$\left. + \frac{2\pi c_P b_P J}{3\alpha}\left(Y_{10}^{*2}(\vartheta_P, \varphi_P) + Y_{11}^*(\vartheta_P, \varphi_P)Y_{1-1}^*(\vartheta_P, \varphi_P)\right) \right\},$$

$$v_{nm}^{\pm} = -\frac{2\sqrt{2}\pi c_P b_P J}{3\alpha} Y_{10}^*(\vartheta_P,\varphi_P) Y_{1\pm1}^*(\vartheta_P,\varphi_P) \frac{n+1}{2n-1} \sqrt{\frac{(n\mp m-2)(n\mp m-1)(n^2-m^2)}{(2n+1)(2n-3)}},$$

$$v_{nm}^{\pm\pm} = \left(\frac{\sigma}{4} + \frac{\pi c_P b_P J}{3\alpha} Y_{1\pm1}^{*2}(\vartheta_P,\varphi_P)\right) \frac{n+1}{2n-1} \sqrt{\frac{(n\mp m-3)(n\mp m-2)(n\mp m-1)(n\mp m)}{(2n+1)(2n-3)}}.$$

Here

$$b_P = \frac{4P^{3/2}}{3(1+P)^3 - 16P^{3/2}}, \quad c_P = \frac{(1+P)^3}{3(1+P)^3 - 16P^{3/2}}$$

are model-dependent coefficients determined by the spin polarization factor $P$ ($0 < P < 1$), the dimensionless spin-polarized current parameter $J$ is defined as

$$J = \frac{v\mu_0 M_s^2 J_e}{kT J_p} \tag{19}$$

where $J_e$ is the current density, taken as positive when the electrons flow from the free into the fixed layer, and $J_p = \mu_0 M_s^2 |e| d / \hbar$ ($e$ is the electronic charge, $\hbar$ is Planck's reduced constant, and $d$ is the thickness of the free layer). A typical value of $J_p$ for a 3 nanometer thick layer of cobalt is $J_p \approx 1.1 \cdot 10^9$ A/cm² while the largest current density reported in experiments is $J_e \approx 10^7 - 10^8$ A cm⁻² (cf. Ref. [20], p. 237). However, for weak damping $\alpha \ll 1$, the ratio $c_P b_P J / \alpha$ appearing in the coefficients $x_{nm}^0$, etc. listed above may be of the same order of magnitude as the anisotropy parameters $\sigma$ and $\delta$ so explaining the strong STT effects on the magnetization dynamics. In contrast, for high damping $\alpha \geq 1$, the STT effects become very small [28].

## APPENDIX B : CALCULATION OF THE STATISTICAL MOMENTS VIA MATRIX CONTINUED FRACTIONS

The differential-recurrence relations Eq. (16) can be solved by matrix continued fraction methods just as in the case of zero STT term [41] with some modifications of the algorithm. By introducing vectors $\mathbf{c}_n^k(\omega)$ $(n = 0, 1, 2, ...)$ with elements composed of the Fourier amplitudes $c_{nm}^k(\omega)$ in Eq. (10), viz.,

$$\mathbf{c}_0(\omega) = \left(c_{00}^0\right), \quad \mathbf{c}_n^k(\omega) = \begin{pmatrix} c_{2n-2n}^k(\omega) \\ \vdots \\ c_{2n2n}^k(\omega) \\ c_{2n-1-2n+1}^k(\omega) \\ \vdots \\ c_{2n-12n-1}^k(\omega) \end{pmatrix},$$

we have from Eq. (16) a matrix differential-recurrence relation for the $\mathbf{c}_n^k(\omega)$, viz.,

$$\mathbf{q}_n^- \mathbf{c}_{n-1}^k(\omega) + \left(\mathbf{q}_n - ik\tau_N\omega\mathbf{I}\right)\mathbf{c}_n^k(\omega) + \mathbf{q}_n^+ \mathbf{c}_{n+1}^k(\omega)$$
$$+\mathbf{p}_n^- \left[\mathbf{c}_{n-1}^{k-1}(\omega) + \mathbf{c}_{n-1}^{k+1}(\omega)\right] + \mathbf{p}_n \left[\mathbf{c}_n^{k-1}(\omega) + \mathbf{c}_n^{k+1}(\omega)\right] \quad (20)$$
$$+\mathbf{p}_n^+ \left[\mathbf{c}_{n+1}^{k-1}(\omega) + \mathbf{c}_{n+1}^{k+1}(\omega)\right] = 0,$$

where the supermatrixes $\mathbf{q}_n^-, \mathbf{q}_n, \mathbf{q}_n^+, \mathbf{p}_n^-, \mathbf{p}_n, \mathbf{p}_n^+, \mathbf{r}_n^-, \mathbf{r}_n, \mathbf{r}_n^+$ are (cf. Eq. (C3) of [28])

$$\mathbf{q}_n = \begin{pmatrix} \mathbf{X}_{2n}^0 & \mathbf{W}_{2n}^0 \\ \mathbf{Y}_{2n-1}^0 & \mathbf{X}_{2n-1}^0 \end{pmatrix}, \mathbf{q}_n^+ = \begin{pmatrix} \mathbf{Z}_{2n} & \mathbf{Y}_{2n}^0 \\ \mathbf{0} & \mathbf{Z}_{2n-1} \end{pmatrix}, \mathbf{q}_n^- = \begin{pmatrix} \mathbf{V}_{2n} & \mathbf{0} \\ \mathbf{W}_{2n-1}^0 & \mathbf{V}_{2n-1} \end{pmatrix},$$

$$\mathbf{p}_n = \begin{pmatrix} \mathbf{X}_{2n}^1 & \mathbf{W}_{2n}^1 \\ \mathbf{Y}_{2n-1}^1 & \mathbf{X}_{2n-1}^1 \end{pmatrix}, \mathbf{p}_n^+ = \begin{pmatrix} \mathbf{0} & \mathbf{Y}_{2n}^1 \\ \mathbf{0} & \mathbf{0} \end{pmatrix}, \mathbf{p}_n^- = \begin{pmatrix} \mathbf{0} & \mathbf{0} \\ \mathbf{W}_{2n-1}^1 & \mathbf{0} \end{pmatrix}.$$

Here the submatrices $\mathbf{V}_n$, $\mathbf{Z}_n$, $\mathbf{W}_n^i$, $\mathbf{X}_n^i$, and $\mathbf{Y}_n^i$ ($i = 0, 1$) have virtually the same form and the same nonzero elements as the submatrices $\mathbf{V}_n$, $\mathbf{W}_n$, $\mathbf{X}_n$, $\mathbf{Y}_n$, and $\mathbf{Z}_n$ from Ref. [28] defined in terms of the *time-independent* elements $v_{nm}$, $w_{nm}$, etc. The only differences which occur are highlighted by the superscript $i = 0, 1$ in the submatrices $\mathbf{W}_n^i$, $\mathbf{X}_n^i$, and $\mathbf{Y}_n^i$, indicating that the elements $w_{nm}$, $x_{nm}$, $y_{nm}$, etc. appearing in these submatrices must now be replaced by $w_{nm}^i$, $x_{nm}^i$, $y_{nm}^i$, etc., respectively.

Next, we introduce super column vectors via

$$\mathbf{C}_0 = \begin{pmatrix} \vdots \\ 0 \\ 0 \\ \mathbf{c}_0^0 \\ 0 \\ 0 \\ \vdots \end{pmatrix}, \quad \mathbf{C}_n = \begin{pmatrix} \vdots \\ \mathbf{c}_n^{-2}(\omega) \\ \mathbf{c}_n^{-1}(\omega) \\ \mathbf{c}_n^0(\omega) \\ \mathbf{c}_n^1(\omega) \\ \mathbf{c}_n^2(\omega) \\ \vdots \end{pmatrix} \quad n = 1,2,3,\ldots, \quad (21)$$

then we have from Eq. (20) the tridiagonal matrix recurrence relations

$$\mathbf{Q}_1 \mathbf{C}_1 + \mathbf{Q}_1^+ \mathbf{C}_2 = -\mathbf{Q}_1^- \mathbf{C}_0, \quad (22)$$

$$\mathbf{Q}_n \mathbf{C}_n + \mathbf{Q}_n^+ \mathbf{C}_{n+1} + \mathbf{Q}_n^- \mathbf{C}_{n-1} = 0. \quad (23)$$

Here $n = 1, 2, 3, \ldots$ and the tridiagonal supermatrices $\mathbf{Q}_n$ and $\mathbf{Q}_n^\pm$, and column vectors $\mathbf{Q}_1^- \mathbf{C}_0$ and $\mathbf{C}_n$ are defined as

$$\left[\mathbf{Q}_n^\pm\right]_{lm} = \delta_{l-1m}\mathbf{p}_n^\pm + \delta_{lm}\mathbf{q}_n^\pm + \delta_{l+1m}\mathbf{p}_n^\pm,$$

$$\left[\mathbf{Q}_n\right]_{lm} = \delta_{l-1m}\mathbf{p}_n + \delta_{lm}\left(\mathbf{q}_n - im\tau_N\omega\mathbf{I}\right) + \delta_{l+1m}\mathbf{p}_n,$$

$$\mathbf{Q}_1^-\mathbf{C}_0 = \frac{1}{\sqrt{4\pi}}\begin{pmatrix} \vdots \\ 0 \\ 0 \\ \mathbf{p}_1^- \\ \mathbf{q}_1^- \\ \mathbf{p}_1^- \\ 0 \\ 0 \\ \vdots \end{pmatrix}, \quad \mathbf{p}_1^- = \begin{pmatrix} 0 \\ 0 \\ 0 \\ 0 \\ 0 \\ w_{1-1}^{+1} \\ w_{10}^{1} \\ w_{11}^{-1} \end{pmatrix}, \quad \mathbf{q}_1^- = \begin{pmatrix} v_{2-2}^{++0} \\ v_{2-1}^{+0} \\ v_{20}^{0} \\ v_{21}^{-0} \\ v_{22}^{--0} \\ w_{1-1}^{+0} \\ w_{10}^{0} \\ w_{11}^{-0} \end{pmatrix},$$

The exact solution of Eqs. (22) and (23) is then rendered by the matrix continued fraction

$$\mathbf{C}_1 = \mathbf{S}_1\mathbf{Q}_1^-\mathbf{C}_0, \qquad (24)$$

where $\mathbf{S}_1$ is defined by the recurrence equation

$$\mathbf{S}_n = -\left[\mathbf{Q}_n + \mathbf{Q}_n^+\mathbf{S}_{n+1}\mathbf{Q}_{n+1}^-\right]^{-1}.$$

The vector $\mathbf{C}_1$ in Eq. (24) contains all the Fourier amplitudes needed for both the linear and nonlinear ac stationary responses. These results are valid for arbitrary field strength meaning that calculating the $c_{nm}^k(\omega)$ and thus the forced response may be reduced to computing matrix continued fractions. When the spin-polarized current parameter $J = 0$, i.e., omitting STT, the above solution agrees in all respects with that given in Ref. [41] for the ac response of a nanomagnet with biaxial anisotropy subjected to superimposed external ac and dc fields of arbitrary strength and orientation.

The solution Eq. (24) is easily computed on a standard PC as the matrix continued fraction involved converge rapidly in most cases. In our calculation, the infinite matrix continued fraction $\mathbf{S}_1$ was approximated by (i) a matrix-continued fraction of finite order (by putting $\mathbf{S}_{n+1} = \mathbf{0}$ at some $n = N$) and by (ii) a finite dimension of the vector $\mathbf{C}_1$ (by choosing the number of harmonics $k = K$) in such way that a further increase of $N$ and $K$ did not change the significant digits in calculated observables. The values of $N$ and $K$ depend on the model parameters values and must be chosen taking into account the desired degree of accuracy of the calculation. The calculations have shown that the results are stable for the ranges of parameters $0 \leq \sigma < 25$, $0 \leq \xi_0 < 25$, $0 \leq \xi < 10$, and $\alpha \geq 0.005$ yielding an accuracy of not less than six significant digits in most cases. For very low damping ($\alpha < 0.002$), very strong ac fields ($\xi \geq 15$) and very high potential barriers, however, the method is difficult to apply because the matrixes involved become ill conditioned so that numerical inversions might be no longer possible. The problem of convergence of matrix continued fractions is discussed in detail by Risken [29].